\documentclass[aps,prb,preprint,superscriptaddress]{revtex4-1}

\usepackage{graphicx}
\usepackage{epstopdf}
\usepackage{physics}
\usepackage{lipsum}
\usepackage{esint}
\usepackage{mathtools}
\usepackage[T1]{fontenc}

\begin{document}


\title{Topological Superconducting Phase in High-\textbf{\textit{T$_c$}} \\ Superconductor MgB$_{2}$
with Dirac-Nodal-Line Fermions}


\author{Kyung-Hwan Jin}
\affiliation{Department of Materials Science and Engineering, University of Utah, Salt Lake City, UT 84112, United States}
\author{Huaqing Huang}
\affiliation{Department of Materials Science and Engineering, University of Utah, Salt Lake City, UT 84112, United States}
\author{Jia-Wei Mei}
\affiliation{Department of Materials Science and Engineering, University of Utah, Salt Lake City, UT 84112, United States}
\affiliation{Institute for Quantum Science and Engineering, and Department of Physics, Southern University of Science and Technology, Shenzhen, 518055, China}
\author{Zheng Liu}
\affiliation{Institute for Advanced Study, Tsinghua University, Beijing 100084, China}
\author{Lih-King Lim}
\affiliation{Institute for Advanced Study, Tsinghua University, Beijing 100084, China}
\author{Feng Liu}
\email[Corresponding author: ]{fliu@eng.utah.edu}
\affiliation{Department of Materials Science and Engineering, University of Utah, Salt Lake City, UT 84112, United States}
\affiliation{Collaborative Innovation Center of Quantum Matter, Beijing 100084, China}


\date{\today}

\begin{abstract}
Topological superconductors are an intriguing and elusive quantum phase, characterized by topologically protected gapless surface/edge states residing in a bulk superconducting gap, which hosts Majorana fermions. Unfortunately, all currently known topological superconductors have a very low transition temperature, limiting experimental measurements of Majorana fermions. Here we discover the existence of a topological Dirac-nodal-line state in a well-known conventional high-temperature superconductor, MgB$_{2}$. First-principles calculations show that the Dirac-nodal-line structure exhibits a unique one-dimensional dispersive Dirac-nodal-line, protected by both spatial inversion and time-reversal symmetry, which connects the electron and hole Dirac states. Most importantly, we show that the topological superconducting phase can be realized with a conventional $s$-wave superconducting gap, evidenced by the topological edge mode of the MgB$_{2}$ thin films showing chiral edge states. Our discovery may enable the experimental measurement of Majorana fermions at high temperature.
\end{abstract}


\maketitle
\textbf{Introduction}\\
Superconducting and topological states are among the most fascinating quantum phenomena in nature. The entanglement of these two states in a solid material into a topological superconducting state will give rise to even more exotic quantum phenomena, such as Majorana fermions. Recently, much effort has been devoted to searching for topological superconductors (TSCs). The first way to realize a TSC phase is by proximity effect via formation of a heterojunction between a topological material and a superconductor (SC) \cite{Fu,Akhmerov,WangMX}. Cooper pairs can tunnel into a topological surface state (TSS), forming a localized state that hosts Majorana bound states at magnetic vortices \cite{Fu,Akhmerov,WangMX}, or into a spin-polarized TSS leading to half-integer quantized conductance \cite{Stern}. Secondly, TSCs can be made by realizing superconductivity in a topological material \cite{Hor,Wray,Sasaki,Sato,ZhangJ,Kirshenbaum,Zhao} or conversely by identifying topological phase in a superconductor \cite{Sakano,Guan,Lv,WangZF}. Broadly speaking, it is preferred to work with one single material, because interfacing two materials may suffer from interface reaction and lattice mismatch between a SC and a topological material. Regardless of which approach to create a TSC, however, a common challenge is that all the known TSCs to date has a very low transition temperature. For example, up to now, the few superconducting/topological heterostructures are realized with low critical temperature ($T_{c}$) $\sim$ 4 K \cite{WangMX}. The superconducting transition temperature induced by doping and/or pressurizing a few topological insulators has a $T_{c}$ in the range of 4 $\sim$ 9 K \cite{Hor,Wray,Sasaki,Sato,ZhangJ,Kirshenbaum,Zhao}, while the $T_{c}$ for some materials where topological and superconducting phases coexist is $<$ 7K \cite{Sakano,Guan,Lv}. High-temperature two-dimensional (2D) SC FeSe on SrTiO$_{3}$ substrate has recently been shown to also host a 2D topological insulating phase with hole doping, whereas its superconducting phases requires electron doping \cite{WangZF}.

Here, we discover the existence of a topological phase in a conventional SC of MgB$_{2}$ with a $T_{c}$ of $\sim$40 K, the highest transition temperature known for a bulk BCS SC. Based on first-principles calculations, we demonstrate a topological Dirac nodal line (DNL) \cite{Burkov,Kim,Yu} structure in MgB$_{2}$, exhibiting a unique combination of topological and superconducting properties. The characteristic 1D dispersive DNL is shown to be protected by both spatial-inversion and time-reversal symmetry, which connects the electron and hole Dirac states. Topological surface band of (010) surface of MgB$_{2}$ shows a highly anisotropic band dispersion, crossing the Fermi level within the superconducting gap. The essential physics of the DNL structure in MgB$_2$ is further analyzed by effective tight-binding (TB) models and the effects of superconducting transition on thin films are also studied. Most importantly, Majorana edge mode, having protected chiral bands crossing zero energy, can be realized.

\textbf{Results and Discussion}\\
MgB$_{2}$ has the AlB$_{2}$-type centrosymmetric crystal structure with the space group P6/$mmm$ (191). As shown in Fig. 1a, it is a layered structure with alternating close-packed trigonal layers of Mg while B layers form a primitive honeycomb lattice like graphene. The optimized lattice constants are $a$ = 3.074 {\AA} and $c$ = 3.518 {\AA}, which agree well with the experimental \cite{Jones,Nagamatsu} and other theoretical results \cite{Kortus,Liu}. The first Brillouin zone (BZ) of bulk MgB$_{2}$ and the projected surface BZ of (010) plane are shown in Fig. 1b.

To reveal the electronic and topological properties of MgB$_{2}$, its electronic band structure is calculated with and without spin-orbit coupling (SOC) (see Supplementary Information). Fig. 1c and 1d show the dominant bonding character of B sheet and the bulk band structure of MgB$_{2}$ with SOC, respectively. The band structure exhibits linear dispersions at both K and H point. The resulting band structure can be easily understood in terms of the B sublattice. The bonding character of the bands in Fig. 1d is indicated by color gradient. Dirac bands are derived from B $p_z$ states ($\pi$-bonding) like the graphene Dirac state and other three bands below Dirac bands are from B $p_{x,y}$ states ($\sigma$-bonding) \cite{Kortus}. Mg $s$ states are pushed up by the B $p_z$ orbitals and donate fully their electrons to the B-derived conduction bands. All bands from B $s$ and $p$ orbitals are highly dispersive, the bands from $\sigma$-bonding are more localized in 2D B sheet (no dispersion along $z$ direction), while the Dirac bands are quite isotropic with high dispersion along $z$ direction. Substantial $k_z$ dispersion of the $p_z$ bands produces a Fermi surface that is approximately mirror reflected with respect to a plane between the $k_z$ =0 [in units of 2$\pi$/$c$] and $k_z$=0.5 planes, with one pocket (electon-like) coming from the antibonding and the other (hole-like) from the bonding $p_z$ band.

Remarkably, the Dirac bands are also dispersive along the K-H directions, so that we have a dispersive Dirac-nodal-line structure [Fig. 1e and 1f]. Fig. 1e shows the band structure along the high-symmetry points that are located in the same plane as $k_z$ is varied. At $k_z$ = 0 there is a hole-doped Dirac band with a Dirac-point energy of 1.87 eV, while at $k_z$ = 0.5, there is an electron-doped Dirac band with a Dirac-point energy of $-$1.91 eV. As $k_z$ increases, the hole-doped Dirac state changes to the electron-doped Dirac state continuously, and the critical point is located at $k_{zc}$ = 0.218 where the carrier changes sign. Usually, SOC may open up gaps at the band crossing points; inversion and/or time-reversal symmetry is insufficient to protect the band crossings. But additional symmetry, like nonsymmorphic symmetry, can protect nodal points or lines. For MgB$_2$, the Dirac nodal line is protected by crystal symmetry and SOC may introduce a small gap leading to a topological insulating phase. However, the SOC strength of B is negligibly small ($\sim$$\mu$eV), even weaker than C and N. Consequently, the topological nodal line of MgB$_2$ survives by a combination of inversion, time-reversal and crystal symmetries, along with weak SOC of B. We note that MgB$_2$ also has a nodal chain structure \cite{Bzdusek} among the conduction or valence bands (see Supplementary Information), but these states are located far from the Fermi level.

We define two independent topological $Z_2$ indices denoted by $\zeta_1$ and $\zeta_2$ \cite{Fang}, one on a closed loop wrapping around the nodal line and the other on a cylinder enclosing the whole line, respectively (Fig. S2). The expression of $\zeta_1$ is described by the Berry phase, $\phi=\oint_{C}d\vec{k} \cdot\vec{A}(\vec{k})$,  as a line integral along a closed path,
\begin{equation}
\zeta_1=\frac{1}{\pi}\oint_{C}d\vec{k} \cdot\vec{A}(\vec{k}),
\end{equation}
where $\vec{A}(\vec{k})$ is the Berry connection, $\vec{A}(\vec{k})=i\expval{\nabla}{n}$ . The simplest Hamiltonian near the K(K$^\prime$) point for the Dirac nodal line of MgB$_2$ is
\begin{equation}
h(\vec{k})=v_{F}(k_{x}\sigma_{x}+k_{y}\sigma_{y})+\alpha{\cos}(k_z)\sigma_{0},
\end{equation}
where $\sigma_i$ are Pauli matrices. For any path in the ($k_x$,$k_y$)-plane that goes around the Dirac line, the index $\zeta_1$ is 1, which means the Dirac lines are stable against perturbations. For the topological invariant of the Dirac nodal line, one can calculate the index $\zeta_2$  using the flow of Wannier charge centers on a set of loops covering the enclosing manifold \cite{Bzdusek2,Fang2}. For the DNL in MgB$_2$, the index $\zeta_2$ is 0. But the nodal lines cannot shrink to a point, instead, they appear in pairs and can only be annihilated in pairs \cite{Hyart}. Furthermore, the parities of energy states can be used to assign the $Z_2$ topological invariants in topological Dirac nodal line semimetals. We found that MgB$_2$ is characterized by weak $Z_2$ indices\cite{Fu2} (see Table S1).

To further reveal the topological nature of the Dirac-nodal-line-semimetal (DNLS) state in MgB$_2$, we also calculate the surface states. The existence of TSSs and Fermi arcs is one of the most important signatures of DNLS. Fig. 2 shows the calculated surface-state spectrum of Mg-terminated (010) MgB$_2$ surface using an iterative Green's function method. We calculated the band dispersions perpendicular to the $\widetilde{\Gamma}$-$\widetilde{\text{Z}}$  direction with four representative cuts at different $k_{sy}$ values in the surface BZ, whose momentum locations are indicated in Fig. 1b. As the bulk BZ is projected onto the (010) surface BZ, one can expect that the nodal line is located at $k_{sx}$ = 2/3 along the symmetry line between  $\widetilde{\Gamma}$($\widetilde{\text{Z}}$) and $\widetilde{\text{X}}$($\widetilde{\text{U}}$). The TSSs of DNLS connect two gapless Dirac points, which are the surface projections of the nodal points in the nodal line [Fig. 2a]. Interestingly, as $k_{sy}$ values increase, the type of bulk Dirac band is changed from the hole-doped to electron-doped. Accordingly, the TSSs connecting two Dirac points are also changed from hole to electron type. Moreover, these TSSs are quite anisotropic: the surface band dispersion is almost flat along the $k_{sx}$ direction, but highly dispersive along the $k_{sy}$ direction [Fig. 2a-c]. The B-terminated surface states are shown in Fig. S3.

We now discuss the Fermi surface. The evolution of the Fermi surface is also obtained from the Green's function method for different values of Fermi energy ($E_F$) [Fig. 2d]. The Fermi surface is composed of one Fermi arc, touching at two singularity points where the surface projections of bulk Dirac points locate. The touching points which are indicated by red dots are also the jointed points between hole (h) and electron (e) pockets. As $E_F$ increases, the touching point is shifted from $k_{sy}$=0.5 [in units of 2$\pi$/$c$] to $k_{sy}$=0. Notably we can observe the Fermi arc as long as $E_F$ is between $-$1.91 eV and 1.87eV. Such a large energy window for observing Fermi arc is a unique and useful feature of dispersive DNL in MgB$_2$. Interestingly, beyond the electronic DNL state in MgB$_2$, phononic Weyl nodal lines and their non-trivial phononic arc states was also predicted in MgB$_2$ \cite{Xie}. 

The essential physics of the DNL structure in MgB$_2$ is further characterized by an effective TB model using B $p_z$ orbital (see Supplementary Information) which gives a quite good description with density-functional theory (DFT) calculations [Fig. 3a]. Given the DNL structure of MgB$_2$, interesting features of quantum oscillation (the Shubnikov-de Hass or the de Haas-van Alphen effect) can be observed, to manifest the predicted nontrivial Berry phase \cite{Kim}. In graphene, such non-zero Berry phase has been confirmed \cite{Novoselov,ZhangY}. Similarly, the effective model of MgB$_2$ exhibits a pseudospin vortex texture arising from the band-degeneracy point [Fig. 3b]. We note that the pseudospin is actually independent of $k_z$. One can clearly see the two inequivalent BZ corners (K and K$^\prime$) having different topologies and being characterized with opposite Berry phases. To measure the non-trivial Berry phase via Landau-fan-type analysis, the electron orbits should enclose the vortex points in a magnetic field. Fig. 3c shows the calculated Fermi surface at different energy levels. For an un-doped system, the electron and hole extremum orbits in a magnetic field along $z$-direction at $E_F$ enclose the $\Gamma$ point, giving rise to a 0 Berry phase. However, for a doped system, the Fermi surface topology changes for different doping levels. At $E_F$=+1 eV ($E_F$=$-$1.2 eV), the hole (electron) extremum orbit encloses the vortex point to enable a non-zero Berry phase [Fig. 3b-c]. There are three different (energy) ranges of doping characterized with different Berry phases. If $E_F$ locates in 0.29 eV$<$ $E_F$ $<$3.25 eV ($-$3.48 eV$<$ $E_F$ $<$$-$1.17 eV), the hole (electron) extremum orbit encloses the vortex point for a non-zero Berry phase. For $-$1.17 eV$<$ $E_F$ $<$0.29 eV, the electron (hole) extremum orbit excludes the vortex point having the zero Berry phase. Thus, to observe the non-zero Berry phase, electron or hole doping is needed. It is known that for MgB$_2$, the Mg atoms can be substituted by Al to form Mg$_{1-x}$Al$_{x}$B$_{2}$, and B can be substituted by C to form Mg(B$_{1-y}$C$_{y}$)$_{2}$ \cite{Pena,Wilke,Kortus2}, to achieve electron doping. Using the rigid band and virtual crystal approximation, we have calculated the alloy band structures, which confirm an upward shifting of $E_F$ due to electron doping (Fig. S4). The non-trivial Berry phase can be measured with a doping concentration of x$>$0.17 or y$>$0.07. Moreover, the biaxial compressive strain also induces electron doping in MgB$_2$ (Fig. S5).

On the one hand, our discovery of topological state in MgB$_2$ might not appear surprising from the theoretical point of view, because of its similarity (B hexagonal plane) to graphene/graphite. One the other hand, it is quite surprising from the experimental perspective, considering the fact that MgB$_2$ has been studied for over decades but none of the experiments has detected any topological signature. This is because there are some fundamental differences between MgB$_2$ and graphene/graphite based systems as we have revealed here. In particular, we show that in order to detect the topological signatures in MgB$_2$, one has to do experiments differently from before, e.g. by measuring angle-resolved photoemission spectroscopy on the (001) instead of commonly used (111) surface and magnetoresistance in the doped sample instead of intrinsic one. However, our most important discovery is possibly the topological superconducting state in MgB$_2$ as we discuss below.

We now consider the superconducting effect on DNLS. MgB$_2$ is a conventional BCS superconductor with two superconducting gaps, $\Delta_{\sigma}$ and $\Delta_{\pi}$, which arise from the $\sigma$ and $\pi$ bands of the B electrons, respectively. The magnitude of the energy gap ranges from 1.5 to 3.5 meV for $\pi$ band and 5.5 to 8 meV for $\sigma$ band \cite{Szabo,Giubileo,Choi,Eskildsen,Souma,Xi,Iavarone}. The strong electron-phonon coupling is known in MgB$_2$ as reflected by strong electron-pair formation of the $\sigma$-bonding states. Because the charge distribution of the $\sigma$-bonding states is not symmetrical with respect to the in-plane positions of boron atoms, the $\sigma$-bonding states couple very strongly to the in-plane vibration of boron atoms  \cite{Choi}. On the other hand, the $\pi$-bonding states, which is related with DNLs, form weaker pairs. However, this pairing is enhanced by the coupling to the $\sigma$-bonding states. Due to the coupling between $\pi$ and $\sigma$ states, $\pi$ and $\sigma$ gaps vanish at the same transition temperature $T_{c}$$\sim$39K, although their values are greatly different at low temperatures. Since the DNL in MgB$_2$ originates from the $\pi$ band, we first consider the superconducting gap for $\pi$ band. But the promising TSC usually requires a spin-triplet (odd-parity) paring state. For the bulk DNL with superconductivity, there is no spectral density in the $s$-wave superconducting gap [Fig. S6]. In general, it is hard to realize the TSC phase using bulk state. 

To realize the TSC state, we now focused on the MgB$_2$ thin films. It was predicted that the MgB$_2$ thin films are mechanically stable and could be grown owing to self-doping effect \cite{Tang}. Furthermore, few-monolayer MgB$_2$ has already been synthesized experimentally on some substrates \cite{Cepek,Petaccia,Bekaert1,Cheng}. Fig. 4 shows the band structure of MgB$_2$ (001) thin films. The dispersive DNL structure is projected onto $\overline{\text K}$ point and presented a formation of multiple Dirac states. These apparent Dirac states are also present in the surfaces bands for the two types of surface. The $\pi$ states of boron layer at B-terminated (Mg-terminated) surface are indicated by red (blue) color. The Dirac state of B-terminated surface shows hole type, while Mg-terminated shows electron type. One can see that the Dirac state of B-terminated surface behaves independently. Moreover the $s$-wave and multiple-gap superconductivity in MgB$_2$ thin films is retained up to a high critical temperature of 20-50K \cite{Morshedloo,Bekaert2}. Taking advantage of these unique features of thin films, we can realize 2D TSC state. 

Conventional $s$-wave superconductivity has been utilized to generate TSCs via proximity to some materials \cite{Lutchyn}. The Dirac state in boron layer of B-terminated surface can be considered as the effective 2D hexagonal lattice model. To realize Majorana fermions with the $s$-wave superconducting gap, we consider a hexagonal lattice TSC based on a model of boron layer with Rashba SOC and exchange field. The corresponding TB Hamiltonian is given by

\begin{equation}
\begin{multlined}
H = -t \sum_{\langle i,j \rangle \alpha} c_{i \alpha}^{\dagger} c_{j \alpha}+i\lambda_{R}\sum_{\langle i,j \rangle \alpha\beta} (\boldsymbol{\sigma}_{\alpha \beta} \times \mathbf{d}_{i,j})_{z} \ c_{i \alpha}^{\dagger} c_{j \beta} +V_{z} \sum_{i \alpha} c_{i \alpha}^{\dagger} \sigma_{z} c_{i \alpha} \\
+\Delta_{\pi} \sum_{i}(c_{i \uparrow}^{\dagger}c_{i \downarrow}^{\dagger}+ \textrm{H.c.}) - \mu \sum_{i \alpha} c_{i \alpha}^{\dagger}c_{i \alpha},
\end{multlined}
\end{equation}

where $c_{i \alpha}^{\dagger}$($c_{i \alpha}$) is the creation (annihilation) operator on site $i$ with spin $\alpha$, and $\boldsymbol{\sigma}$ are the Pauli matrices. The $\langle i,j \rangle$ represents the nearest-neighboring (NN) sites. The first term is the NN hopping term in boron layer. The second term is the Rashba SOC arising from a perpendicular electric field to the B layer adjacent to substrate, with $\lambda_{R}$ and $\mathbf{d}_{i,j}$ representing the coupling strength and a unit vector from site $j$ to site $i$, respectively. The $V_{z}$ ($\Delta_{\pi}$) in the third (fourth) term corresponds to the exchange field (superconducting gap of $\pi$ band). The $\mu$ in the last term is chemical potential. We can transform the Hamiltonian of Eq. (3) to the Bogoliubov-de Gennes (BdG) Hamiltonian $H_{\text BdG}$ in the momentum space (see Supplementary Information). 

If neglecting the Rashba and exchange field term, the system is topologically trivial with $s$-wave superconducting gap same as the bulk DNL state. As expected, the superconducting gap mixes the Dirac states, resulting in the disappearance of gapless edge modes. When the Rashba and exchange field are turned on, spin-up and spin-down bands are split and mixed. The 2D BdG Hamiltonian with broken time-reversal symmetry belongs to the topological class D \cite{Schnyder}, which is characterized by an integer number. The Chern number $C_1$ can be calculated by \cite{Ghosh}
\begin{equation}
C_1=\frac{1}{2\pi}\int_{\text BZ}d^2\vec{k} f_{xy}(\vec{k}),
\end{equation}

where $f_{xy}(\vec{k})$ is the Berry curvature

\begin{equation}
 f_{xy}(\vec{k})=i\sum_{m,n}(f_{m}-f_{n})\frac{u_{m\vec{k}}^{\dagger}\partial_{k_x}H_{\text{BdG}}u_{n\vec{k}}u_{n\vec{k}}^{\dagger}\partial_{k_y}H_{\text{BdG}}u_{m\vec{k}}}{(E_{m\vec{k}}-E_{n\vec{k}})^2}.
\end{equation}

Here, $u_{m\vec{k}}$($E_{m\vec{k}}$) is the m-th eigenvector (eigenvalue) of $H_{\text{BdG}}$ ; $f_m$  is Fermi occupation factor. Interestingly, we found that the boron layer of MgB$_2$ thin film is topologically nontrivial in the case of $\lvert V_{z} \rvert \geq \lvert\Delta_{\pi}\lvert$ with the Chern number $C_1$=4. We note that this criterion does not depend on the value of the Rashba coupling. Fig. 5a show the calculated energy spectrum of BdG Hamiltonian near the crossing points with different $V_z$ values. We have chosen the pairing  $\Delta_{\pi}$  as 3 meV, similar to the experimental $s$-wave gap in MgB$_2$. For pristine hexagonal boron layer, the Dirac bands appear at the $\overline{\text K}$  and $\overline{\text K}'$ points, respectively. If the exchange and Rashba SOC interactions are considered, the trivial $s$-wave superconducting gap near the Dirac point turns into a topological gap. As the magnetic exchange field $V_z$ increases, the topological gap increases. Further, we study the topological phase diagram as shown in Fig. 5b. The band gap ($E_g$) is calculated in the ($k_x$,$k_y$)-plane. The phase boundary between the normal SC and TSC is determined by the dashed curves, i.e. $V_{z} =\Delta_{\pi}$. To visualize the formation of TSSs within the superconducting gap, we calculated the Majorana edge states with a semi-infinite boron sheet. Fig. 5c and 5d show the energy spectra of zigzag edge near the $\overline{\text K}$  and $\overline{\text K}'$ points. We found that there exist two zero-energy states in each valley with same propagation direction ($v_F$<0), which is in agreement with $C_1$=4. This indicates that these zero energy states are topologically protected Majorana edge states. Armchair edge also exhibits four protected bands crossing zero energy (Fig. S8).

Finally, we address the experimental feasibility of the TSC from a quantum anomalous Hall (QAH) phase of graphene having a similar structural system with boron layer. Graphene on the magnetic insulator, such as BiFeO$_3$ \cite{Qiao}, RbMnCl$_3$ \cite{ZhangSR}, Cr$_2$Ge$_2$Te$_6$ \cite{ZhangPRB}, can have an exchange field ($V_z$$\sim$70-240 meV) and Rashba SOC ($\lambda_R$$\sim$1-4meV), realizing the QAH state. We suggest that MgB$_2$ thin films be epitaxially grown on these magnetic substrates, and the exchange field and Rashba SOC can be induced in the boron layer adjacent to substrate. Since the MgB$_2$ thin films naturally have $s$-wave superconductivity, the MgB$_2$ thin films are expected to become TSC with the induced exchange field ($V_{z}\geq\Delta_{\pi}$) and Rashba SOC. 

In summary, based on first-principles calculations and model analysis, an intriguing inversion and time-reversal symmetry protected Dirac nodal line state is revealed in high-temperature superconductor MgB$_2$. Our finding provokes an exciting opportunity to study topological superconducting phase in an unprecedented high temperature and may offer a promising material platform to building novel quantum and spintronics devices. It will stimulate future studies of topological phases in a broader range of superconducting materials, such as honeycomb lattice layered structure.

\textbf{Methods}\\
We performed first-principles calculations within the framework of density-functional theory (DFT) using the Perdew-Burke-Ernzerhof-type generalized gradient approximation (GGA) for the exchange-correlation functional, as implemented in the Vienna ab initio simulation package \cite{Kresse,Perdew}. All the calculations are carried out using the kinetic energy cutoff of 500 eV on a 12$\times$12$\times$12 Monkhorst-Pack $k$-point mesh. All structures are fully optimized until the residual forces are less than 0.01 eV/{\AA}. The SOC is included in the self-consistent electronic structure calculation. We construct Wannier representations by projecting the Bloch states from the first-principles calculation of bulk materials onto Mg $s$ and B $s$, $p$ orbitals \cite{Mostofi}. Based on Wannier representations, we further calculate the surface density of states and Fermi surface using the surface Green's function method for the (010) surface of a semi-infinite system \cite{Sancho,Wu}.
\\
\textbf{Data availability}\\
The data that support the findings of this study are available from the corresponding authors upon reasonable request.

\textbf{Acknowledgments}\\
We would like to thank Tom\'a\v{s} Bzdu\v{s}ek, QuanSheng Wu and Alexey A. Soluyanov for helpful discussions. K.-H.J., H.H. and F.L. acknowledge financial support from DOE-BES (No. DE-FG02-04ER46148). We also thank Supercomputing Center at NERSC and CHPC at University of Utah for providing the computing resources. 

\textbf{Author Contributions}\\
K.J. and F.L. designed the research. K.J. performed theoretical calculation, H.H., J.M, Z.L and L.L discussed the results, and K.J and F.L prepared the manuscript.

\textbf{Additional information}\\
Supplementary information is available in the online version of the paper. Reprints and permissions information is available online at www.nature.com/reprints.

\textbf{Competing interests}\\
The authors declare no competing financial interests.


\newpage
 \begin{figure}
 \includegraphics[scale=0.33]{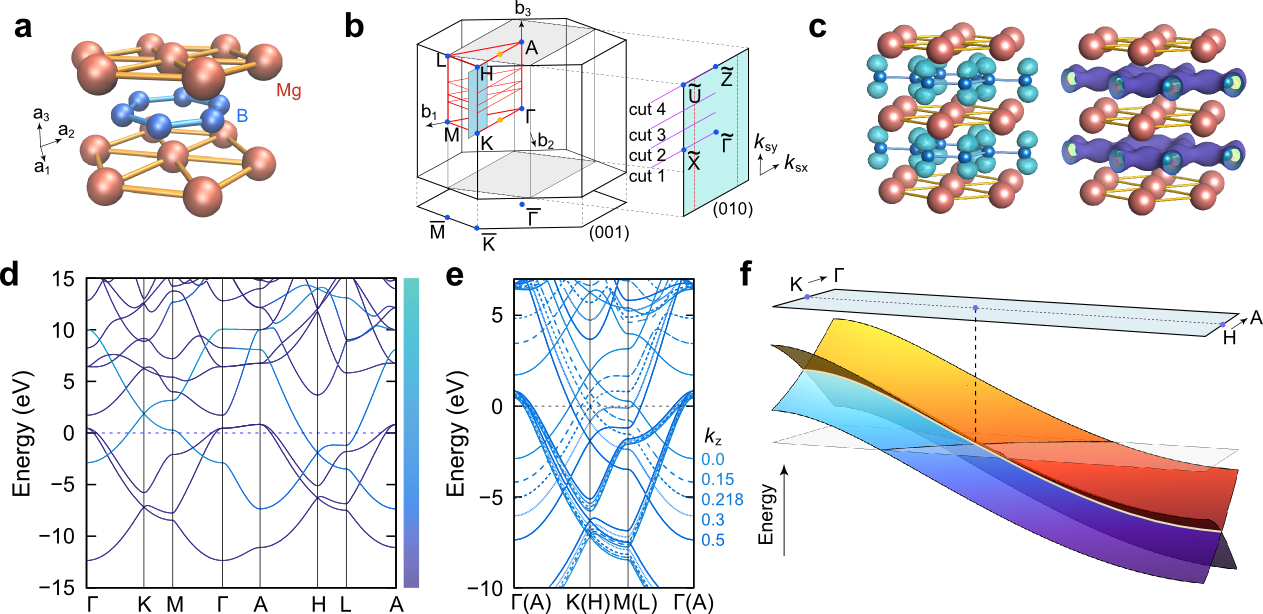}
 \caption{\label{FIG1} Crystal and electronic structure of MgB$_{2}$. {\bf a} Crystal structure of MgB$_{2}$ with P6/$mmm$ symmetry. {\bf b} Bulk BZ and the projected BZ of the (010) and (001) surface with purple lines indicating the locations of cuts 1 to 4 for surface states in Fig. 2{\bf a}. {\bf c} $\pi$ (B $p_z$) and $\sigma$ (B $s$, $p_{x,y}$) bonding character of boron sheet. {\bf d} Band structure of MgB$_{2}$ with the coded bonding character. The color gradient from purple to blue represents varying contribution from $\sigma$ bonding to $\pi$ bonding of boron sheet. {\bf e} Band structure along high symmetry points which are located in the same plane as varying $k_z$ value. {\bf f} Dispersive Dirac nodal line structure along the K-H direction as indicated by a small cyan rectangular in {\bf b}.}
 \end{figure}

\newpage
 \begin{figure}[t]
\centering
\includegraphics[scale=0.33]{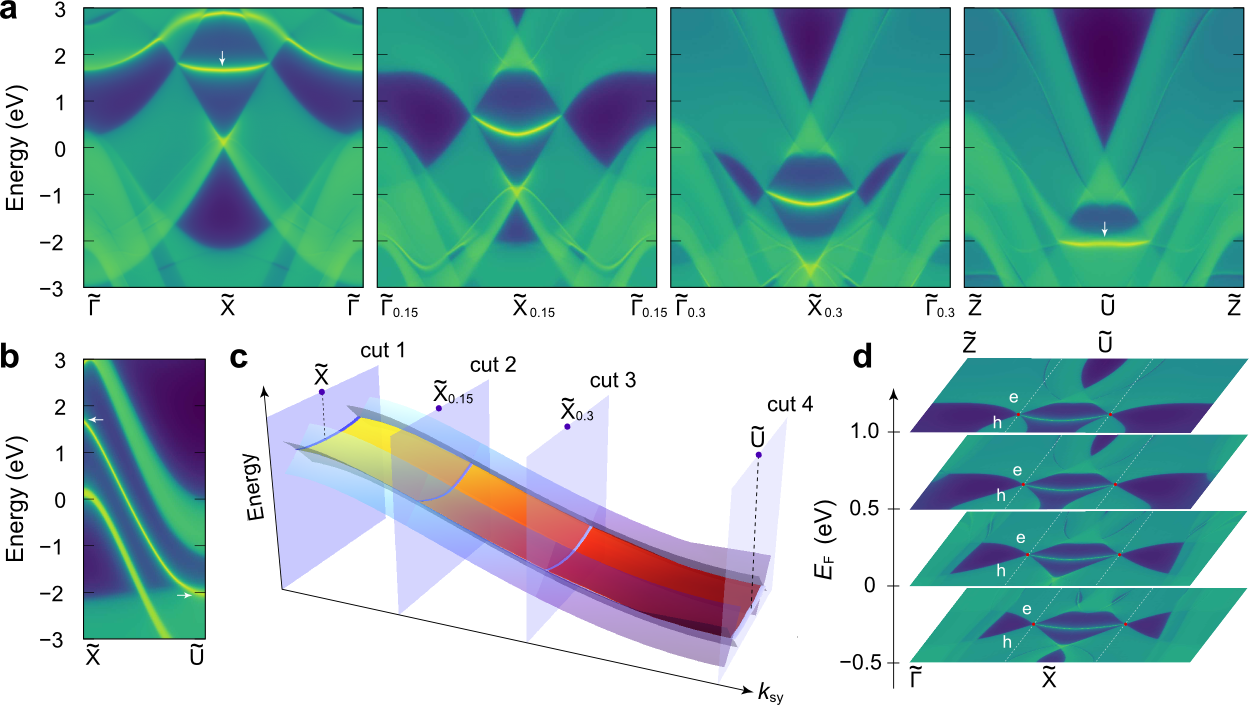}
 \caption{\label{FIG2} Surface states of MgB$_2$ (010) surface. {\bf a} The projected density of state for Mg-terminated (010) surface along the cuts at $k_{sy}$ = 0, 0.15, 0.3, and 0.5 in 2D surface BZ as indicated in Fig. 1{\bf b}, respectively. {\bf b} The surface states along the $\widetilde{\text{X}}$-$\widetilde{\text{U}}$  direction. {\bf c} 3D schematic plot of the electronic surface band structure near two Dirac points. The gray planes indicate the momentum locations of cuts 1 to 4  in {\bf a}. {\bf d} Constant energy contours of Mg-terminated (010) surface at fixed energies $E_F$= $-$0.5, 0, 0.5 and 1.0 eV. The red dots mark two pieces of Fermi arcs connecting surface projection of bulk Dirac points.}
 \end{figure}

\newpage
 \begin{figure}
 \includegraphics[scale=0.33]{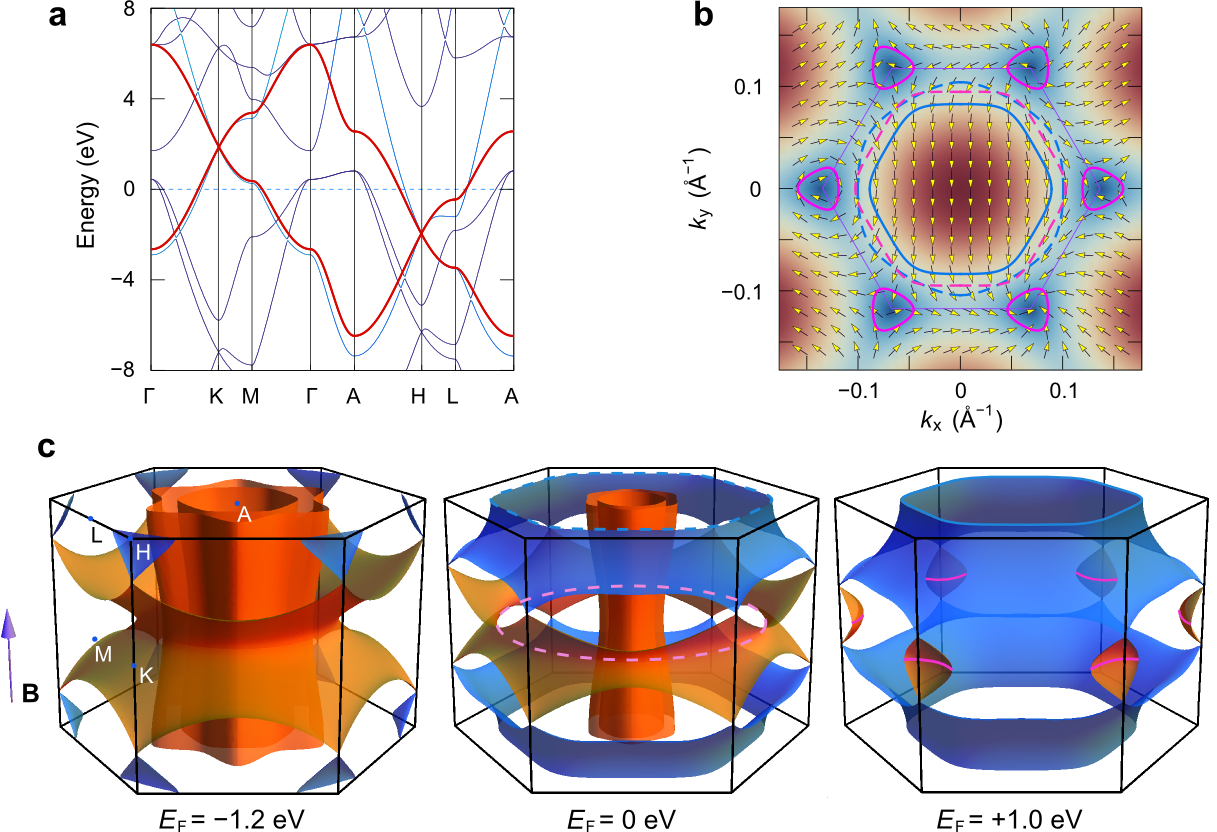}
 \caption{\label{FIG3} Effective model of dispersive DNL of MgB$_2$ and Berry phase analysis. {\bf a} The calculated band structure of effective model (red lines) using B $p_z$ orbitals with parameter $t$=1.50 eV and $t_z$=0.954 eV with DFT band structure. {\bf b} The pseudospin texture with the corresponding valence energies. The arrows represent the pseudospin. The color gradient from red to blue represents energies of valence states. The blue and red dashed circles correspond to the electron and hole extremum orbits in a magnetic field (\textbf{B}) along $z$-direction at $E_F$=0 eV. The blue and red solid circle correspond to the electron and hole extremum orbits at $E_F$=+1.0 eV. {\bf c} The Fermi surfaces for $E_F$=$-$1.2 eV, $E_F$=0 eV and $E_F$=+1.0 eV, respectively. The electron (hole) Fermi surfaces are indicated by blue (red) isosurfaces. The extremum orbits are indicated by dashed line (solid line) for $E_F$=0 eV ($E_F$=+1.0 eV).  }
 \end{figure}

\newpage
 \begin{figure}[t]
 \includegraphics[scale=1.]{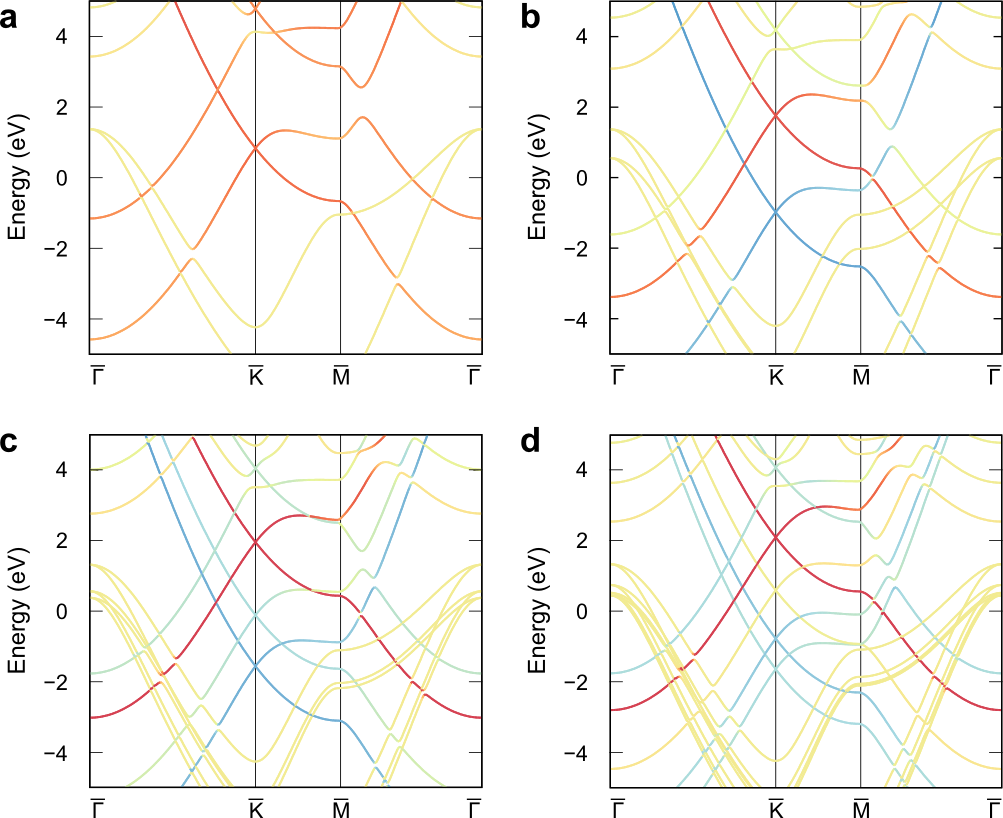}
 \caption{\label{FIG4} Electronic structure of MgB$_2$ (001) thin films. {\bf a}-{\bf d} Calculated band structures of 1-, 2-, 3- and 4-layer of MgB$_2$ (001) thin films, respectively. The red (blue) color represents the $\pi$ state of boron layer at B-terminated (Mg-terminated) surface.}
 \end{figure}

\clearpage
 \begin{figure}[t]
 \includegraphics[scale=1.]{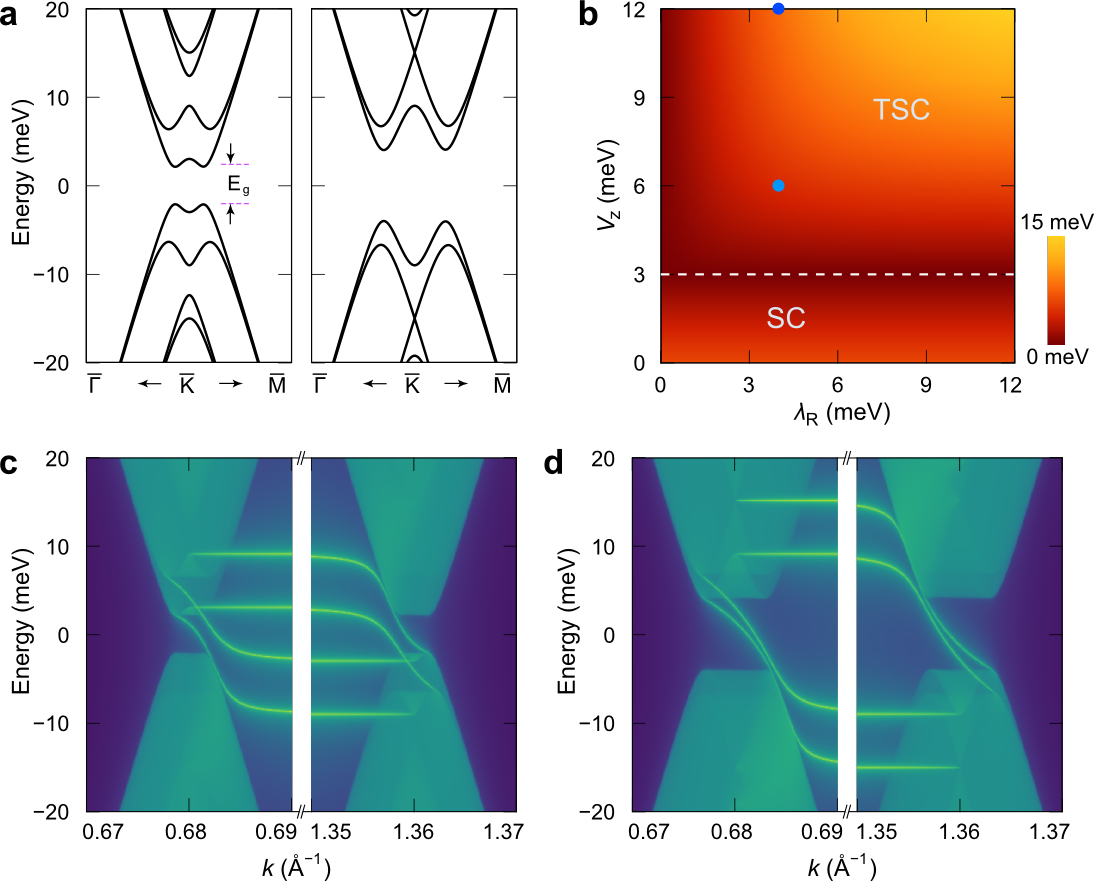}
 \caption{\label{FIG5} The topological edge mode of the MgB$_2$ thin films. {\bf a} The calculated band structure of the BdG Hamiltonian with parameter $t$=1.5 eV, $\mu$=0 eV, $\Delta_{\pi}$ =3 meV, $\lambda_R$ =4 meV, $V_{z}$=6 meV (left) and $V_{z}$=12 meV (right), respectively. {\bf b} Topological phase diagram in the parameter space of $\lambda_R$ and $V_z$. The color indicates the bandgap ($E_g$) in the ($k_x$, $k_y$)-plane at $k_z$=0.25. The band structure parameters for {\bf a} is marked by the dot. The dashed line is the boundary between normal SC and TSC. {\bf c}-{\bf d} The calculated edge spectra for the zigzag edge of the semi-infinite boron sheet with parameters in {\bf a}, respectively.}
 \end{figure}

\end{document}